\documentclass[preprint2]{aastex}

\usepackage{color}
\usepackage{ulem}

\shorttitle{Analysis of detection efficiency for GCs pulsars}
\shortauthors{Yin et al.}

\begin{document}

\title{The Analyses of  Globular Clusters Pulsars and Their Detection Efficiency}

\author{De-Jiang Yin\altaffilmark{1}, Li-Yun Zhang\altaffilmark{1, *}, Bao-Da Li\altaffilmark{1}, Ming-Hui Li\altaffilmark{2}, Lei Qian\altaffilmark{3,4,5,6,*}, Zhichen Pan\altaffilmark{3,4,5,6,*}}

\affil{$^1$College of Physics, Guizhou University, Guiyang 550025, China}
\email{liy\_zhang@hotmail.com, dj.yin@foxmail.com}
\affil{$^2$State Key Laboratory of Public Big Data, Guizhou University, Guiyang 550025, China}
\affil{$^3$National Astronomical Observatories, Chinese Academy of Sciences, 20A Datun Road, Chaoyang District, Beijing 100101, China}
\email{panzc@bao.ac.cn, lqian@nao.cas.cn}
\affil{$^4$State Key Laboratory of Space Weather, Chinese Academy of Sciences,  Beijing  100190, China}
\affil{$^5$CAS Key Laboratory of FAST, National Astronomical Observatories, Chinese Academy of Sciences, Beijing 100101, China}
\affil{$^6$College of Astronomy and Space Sciences, University of Chinese Academy of Sciences, Beijing 100049, China}

\begin{abstract}
  Up to November 2022, 267 pulsars have been discovered in 36 globular clusters (GCs).
  In this paper, we present our studies on the distribution of GC pulsar parameters and the detection efficiency.
  The power law relation between the average of dispersion measure ($\overline{\rm DM}$) and
  dispersion measure difference ($\Delta {\rm DM}$) of known pulsars in GCs is
  $\lg\Delta {\rm DM} \propto 1.52 \lg \overline{\rm DM}$.
  The sensitivity could be the key to find more pulsars.
  As a result, several years after the construction of a radio telescope,
  the GC pulsar number will be increased accordingly.
  We suggest that currently GCs in the southern hemisphere could have higher possibilities to find new pulsars.
\end{abstract}

\keywords{Globular Cluster; Pulsar}

\section{Introduction}
\label{sect:intro}

There have been 170 globular clusters (GCs) in the Milky Way till 2021 \citep{2021MNRAS.505.5978V}.
Since the discovery of the first GC pulsar in M28 \citep{1987Natur.328..399L},
till November 2022,
267 pulsars have been discovered in 36 GCs\footnote{Paulo Freire's GC Pulsar Catalog, https://www3.mpifr-bonn.mpg.de/staff/pfreire/GCpsr.html}.
Terzan 5 and 47 Tucanae harbor the largest GC pulsar populations, with 40 and 29 pulsars, respectively.
Among all GC pulsars,
147 are members of binary systems, while 120 are isolated.
The distance of GCs with known pulsars ranges from 2.2 kpc (M4, 1 pulsar; \citealt{1988Natur.332...45L}) to 17.9 kpc (M53, 5 pulsars, e.g., \citealt{1991Natur.349...47K}).

There are many exotic objects in GC pulsars,
including the fastest spinning pulsar (J1748-2446ad, with the spin frequency $\thicksim$ 716 Hz, \citealt{2006Sci...311.1901H}),
the triple system with a white dwarf and a Jupiter-mass companion in M4 (B1620-26, \citealt{2003Sci...301..193S}),
binaries with highly eccentric orbits (e.g., NGC6652A, with eccentricity $e$ $\thicksim$ 0.968, \citealt{2015ApJ...807L..23D};
NGC 1851A, with $e$ $\thicksim$ 0.888, \citealt{2004ApJ...606L..53F}),
17 redbacks and 26 black widows.
These pulsars can be used to constrain the stellar evolution, and to study the intrinsic dynamics of GCs (e.g., \citealt{2003ApJ...591L.131P, 2013Natur.501..517P}).

By the end of 2004, only 100 pulsars were detected in 24 GCs \citep{2005ASPC..328..147C}.
The GCs with largest numbers of pulsars were Terzan 5 (with 23 pulsars) and 47 Tucanae (with 22 pulsars).
Between 2004 and 2007, 38 more GC pulsars were discovered while the number of GCs with pulsar discoveries only increased slightly, from 24 to 25.
At that moment, the only two factors related to the number of pulsars in a GC are the total mass and the distance.
However, there are exceptions, such as $\omega$ Centauri (NGC 5139),
one of the nearest and most massive GCs in Milky Way \citep{2008AIPC..983..415R}.
In the following decade, the number of GC pulsars increased slowly.

Most of GC pulsars are discovered by large radio telescopes.
These telescopes include
Lovell ( $\thicksim$ 5 pulsars; e.g., \citealt{1987Natur.328..399L}),
Giant Meterwave Radio Telescope (GMRT) ($\thicksim$ 2 pulsars; e.g., \citealt{1991ASPC...19..376S}),
Parkes ($\thicksim$ 48 pulsars; e.g., \citealt{2002MNRAS.335..275M}),
Arecibo ($\thicksim$ 28 pulsars; e.g., \citealt{2007ApJ...670..363H}),
Green Bank Telescope (GBT) ($\thicksim$ 82 pulsars; e.g., \citealt{2009IEEEP..97.1382P}),
MeerKAT ($\thicksim$ 51 pulsars; e.g., \citealt{2022A&A...664A..27R}; TRAPUM project\footnote{http://trapum.org/discoveries/})
and the Five-hundred-meter Aperture Spherical radio Telescope (FAST, \citealt{2011IJMPD..20..989N}) ($\thicksim$ 41 pulsars;  e.g., \citealt{2021ApJ...915L..28P}; FAST GC survey\footnote{https://fast.bao.ac.cn/cms/article/65/}).

In this work, we analyzed the 267 GC pulsars and estimated the detection efficiency.
The detection efficiency of radio telescopes and the spatial distribution of GC pulsars are given in section 2
 and \ref{Spatial distribution}.
The distribution Of Physical parameters are presented in section \ref{DM and delDM}.
Section \ref{sect:conclusion} is for the summary.

\section{The detection efficiency of  large radio telescopes}
\label{Large radio telescopes}

\begin{figure*}
\centering
    \includegraphics[width=10cm]{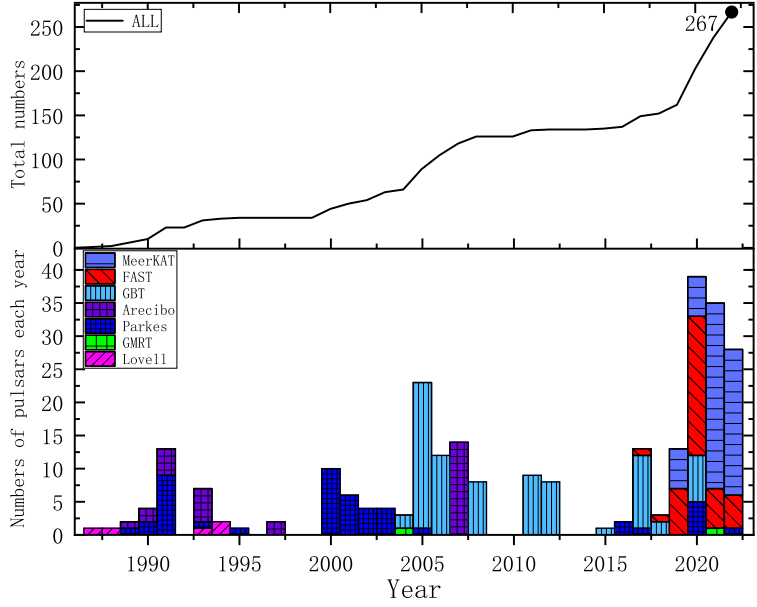}
    \caption{
Cumulative number of known pulsars in GCs.
The upper subplot: All GC pulsars; the lower subplot: the numbers of GC pulsars discovered by different radio telescopes as the colored bars.
For the definition of discovery time:
the discovery date of pulsars from FAST and MeerKAT are listed on their websites,
pulsars discovered by other telescopes are the "publication" time,
based on the time of the first reference listed by Paulo's GC pulsar catalog.}
   \label{Fig01}
\end{figure*}

About 2 years after the construction of a radio telescope,
the number of GC pulsars tend to be increased accordingly as shown in Figure \ref{Fig01}.
For example, FAST and MeerKAT are the main facilities for GCs pulsar discoveries in recent years,
.
The first GC pulsars discovery by FAST were reported in 2020 (e.g., \citealt{2020ApJ...892...43W, 2020ApJ...892L...6P}).
Till now, a total of 41 pulsars have been found.
The MeerKAT started observation in 2018,
and reported the first GC pulsar discovery in 2021 \citep{2021MNRAS.504.1407R}.
In total, 61 pulsars have been detected by MeerKAT.
As a result, the number of GC pulsars increased steeply since 2020.

The possibility to discover GC pulsars is mainly related to the sensitivity and the number of visible clusters within the sky covered by a telescope.
The achievement of higher sensitivity would increase the number of GC pulsar discoveries.
The minimum detectable flux density for a pulsar
can be calculated with the radiometer equation (e.g., \citealt{1985ApJ...294L..25D}):
\begin{equation}
S_{\rm min} = \frac{\alpha\beta T_{\rm sys}}{G\sqrt{n_{\rm pol}\Delta\nu t_{\rm int}}}\sqrt{\frac{W}{P-W}},
\end{equation}
where $\alpha$ is the minimum signal-to-noise ratio (S/N),
$\beta$ is the factor accounting for the losses from the digitization and other processing,
$T_{\rm sys}$ is the system temperature,
$G$ is the gain of the telescope, $n_{\rm pol}$ is the number of polarizations,
$\Delta\nu$ is the observing bandwidth (in the unit of MHz),
$t_{\rm int}$ is the integration time in the unit of seconds,
$W$ is the pulse width, and $P$ is the pulse period.
Here, we take $\alpha$ = 10, $n_{\rm pol}$ = 2, $t_{\rm int}$ = 2 h and $W$ = 0.1$P$.

The data used to calculate the minimum detectable flux density for each radio telescope is shown in Table \ref{tab2}.
Here, we define that the detection efficiency ($\eta$) is the numbers of pulsars found by each radio telescope ($N_{\rm p}$) divided by the numbers of visible GCs in its sky ($N_{\rm gc}$):

\begin{equation}
\eta = N_{\rm p}/N_{\rm gc}.
\end{equation}

\begin{figure*}
\centering
    \includegraphics[width=10cm]{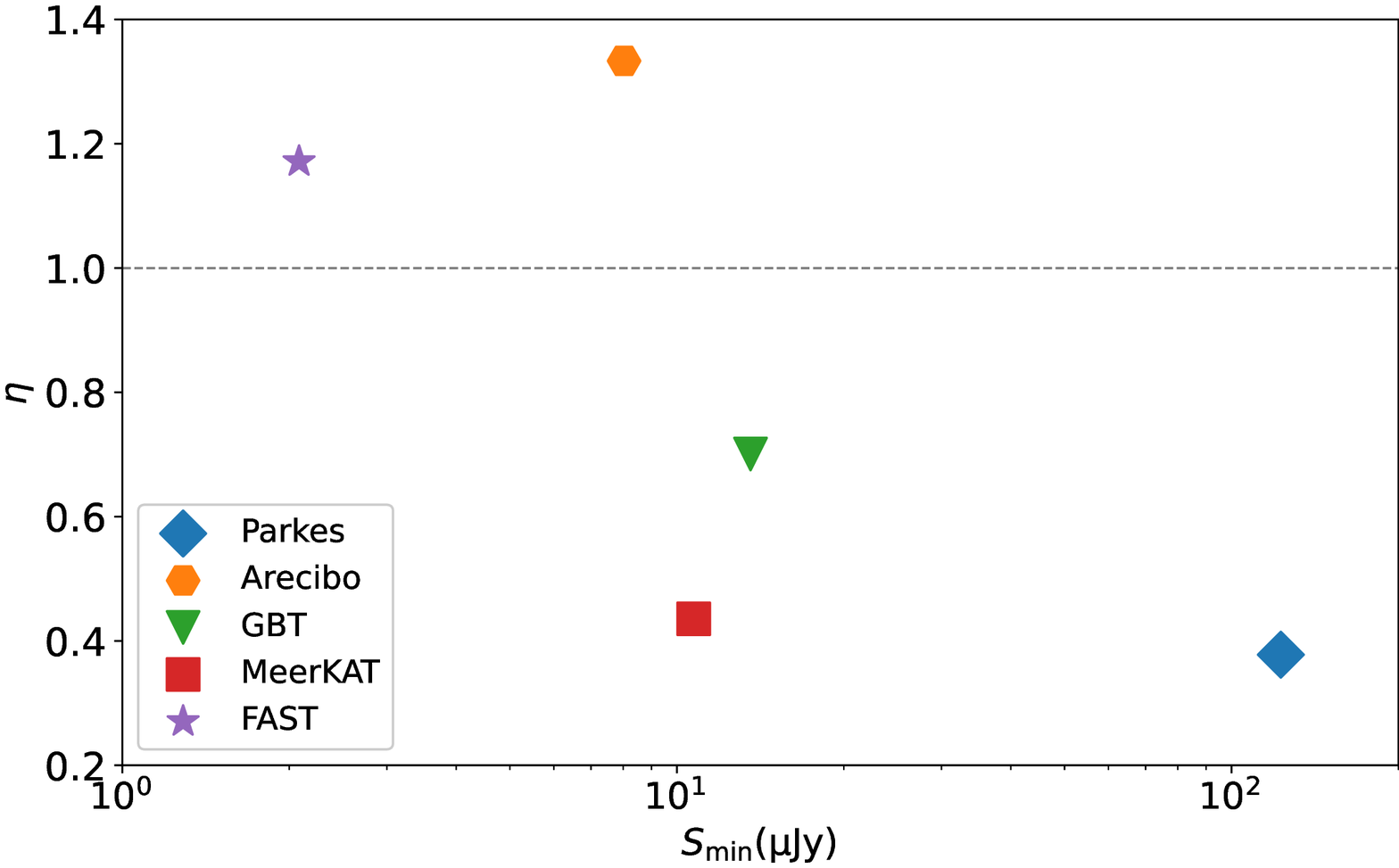}
    \caption{The detection efficiencies of radio telescopes. The uGMRT was not included.}
    \label{Fig02}
\end{figure*}

The most distant GC with known pulsars is M53 (17.9 kpc).
Accordingly,
We use 30 kpc as the limit for detecting GC pulsars for current radio telescopes.
For 157 GCs, 141 are in this range.
Figure \ref{Fig02} shows the relation between the minimum detection flux density and the pulsar detection efficiency of each radio telescope.

\begin{table*}[htpb]
\centering
\caption{parameters of GC pulsar surveys. (1). Parkes: 13 beam receiver package at a central
radio frequency of 1374 MHz \citep{2000ApJ...535..975C};
(2). Arecibo: using the Gregorian L-band wide receiver, the central observing frequency was either 1175 MHz (DM $\lesssim$ 100 pc cm$^{3}$) or 1475 MHz (DM $\gtrsim$ 100 pc cm$^{3}$) \citep{2007ApJ...670..363H};
(3). GBT: the Green Bank Ultimate Pulsar Processing Instrument backend at S band (2 GHz) \citep{2015ApJ...807L..23D};
(4). FAST: The FAST 19-beam receiver with a frequency range of 1.0-1.5 GHz \citep{2021ApJ...915L..28P};
(5). MeerKAT:  used the L-band (856¨C1712 MHz) receivers and up to 42 of the 44 antennas with central frequency 1284 MHz \citep{2021MNRAS.504.1407R},  $G = (N_{\rm ant}/64) \times 2.8 {\ \rm K\ Jy}^{-1}$ and $N_{\rm ant}$ the number of antennas used in the observation \citep{2022MNRAS.513.2292A};
(6). uGMRT: the upgraded Giant Metrewave radio telescope (uGMRT), observed  the low-DM clusters at 400 MHz and the others at 650 MHz with  the digital GMRT wideband backend system.
$T_{\rm sys}$ is 130 K for 400 MHz and 102.5 K for 650 MHz \citep{2022arXiv220515274G}.
We take $\alpha$ = 10, $n_{\rm pol}$ = 2,
 = 2 h and $W$ = 0.1$P$.}
\label{tab2}
\begin{tabular}{ccccccccc}
\hline
\textbf{Telescope} & \textbf{$\beta$} & $T_{\rm sys}$ & \textbf{Gain} & \textbf{Bandwidth} &  \textbf{$S_{\rm min}$} &\textbf{$\eta$} & Visible Sky   &  Visble-GCs \\
              &         &  (K)       &  (K/Jy)              &    (MHz)               &      ($\mu$Jy)        &        &  (Dec/degree)    &  ($R_{\rm Sun}\leq 30 $ kpc)  \\
\hline
uGMRT$^{(6)}$              & 1          & 102.5       & 4.2                 & 200              &        47.9     &       0.01      &   (-71, 90)      &      136     \\

Parkes$^{(1)}$             & 1.5        & 35         & 0.7                 & 288              &      122.8     &           0.38      &   (-87, 10)                &      127   \\
Arecibo$^{(2)}$            & 1.2        & 40         & 10.5                 & 250              &      8.0      &          1.33       &   (-1.5, 38.5)             &       21   \\
GBT$^{(3)}$                & 1.05       & 22.8       & 1.9                 & 700              &      13.6     &         0.70         &   (-46, 90)                 &      117    \\
FAST$^{(4)}$               & 1          & 24        & 16                    & 400              &      2.1     &         1.17         &   (-14, 65)          &       35 \\
MeerKAT$^{(5)}$            & 1.1        & 26        & 2.8                  & 700               &       10.7     &       0.44          &    (-90, 45)             &        140   \\
uGMRT$^{(6)}$              & 1          & 102.5       & 4.2                 & 200              &        47.9     &       0.01      &   (-71, 90)      &      136     \\
\hline
\end{tabular}
\end{table*}

With higher sensitivity, it is possible to detect pulsars missed by previous surveys.
The total discovery rate of pulsars in GCs is $\eta$ = 267/141 $\sim$ 1.89.
This is due to the accumulation of all GC pulsar discoveries.
The detection efficiency with FAST or Arecibo is higher than that of other radio telescopes.
This can be explained by their high sensitivities.
Similarly, Parkes is limited by its relatively low sensitivity (64-meter single dish).
With the UWL \citep{2020PASA...37...12H} installed, it may have chance to find more.
GBT has a large sky coverage and an intermediate sensitivity.
Its detection efficiency is improved mainly due to the 35 pulsars from Terzan 5 (e.g., \citealt{2005Sci...307..892R}).
It is worthy noting that FAST, Arecibo, and GBT locate in the north hemisphere,
while more than half  of the GC pulsars discovered by them are located in the southern sky.
MeerKAT located in South Africa (south hemisphere) and has already discovered a number of GC pulsars.
The number of visible GCs in the FAST and MeerKAT skies is $\sim$ 45 and 140, respectively.
MeerKAT will definitely have more discoveries in the near future.
Assuming that the detection efficiency of MeerKAT is 1,
the total GC pulsars discovered can reach $\sim$ 140.

The next stage is the era of SKA, which will revolutionize the population detection of GCs pulsars.
The SKA will provide unprecedented sensitivity (e.g., $\sim$4 $\mu$Jy in 2-hr integrations operating from 1.4 - 2.0 GHz of SKA1-MID) for GC search of targets in the southern sky and can cover about $\sim$ 154 GCs \citep{2015aska.confE..47H}.
Considering the integration time of 2 hours, we assume that the detection efficiency of SKA is 1, and about 150 GC pulsars will be detected.
For SKA1-MID, tracking observation times of up to 8 hours ($\sim$2 $\mu$Jy) are possible, the SKA possibly detects more GC pulsars (e.g., \citealt{2015aska.confE..47H}).

\section{Spatial distribution of GCs with known pulsars}
\label{Spatial distribution}
\begin{figure*}
\centering
    \includegraphics[width=10cm]{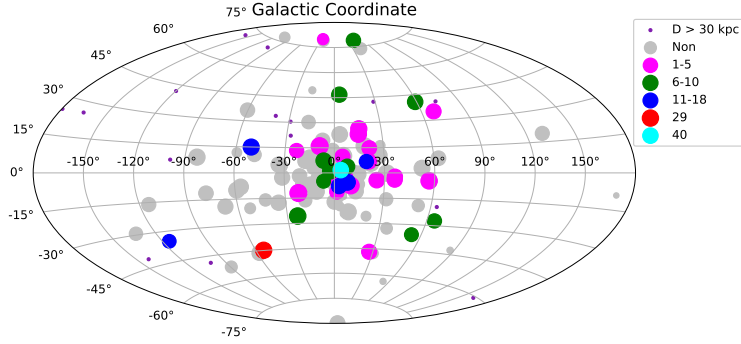}
    \caption{Distribution of GCs in the Galactic coordinate.
            The circles in different colours represent the different numbers of known pulsars in GCs.
            The size of the circles represents the distances from Earth to GCs.
            Larger circles indicate closer distances.
            The same size of circles are for clusters with a distance greater than 30 kpc.
            The distances of GCs are from (\citealt{1996AJ....112.1487H}, 2010 edition).}
    \label{Fig03}
\end{figure*}

The most distant GC pulsars were located in M53,
and 35 GCs are farther away than M53 (\citealt{1996AJ....112.1487H}, 2010 edition).
Figure \ref{Fig03} shows the spatial distribution of all GCs in Galactic coordinate.
GCs with known pulsars do not featured distribution or bias when compared with other GCs.
The details of pulsars in each GC are shown in attached Table \ref{tab1}.

\begin{figure*}
\centering
    \includegraphics[width=10cm]{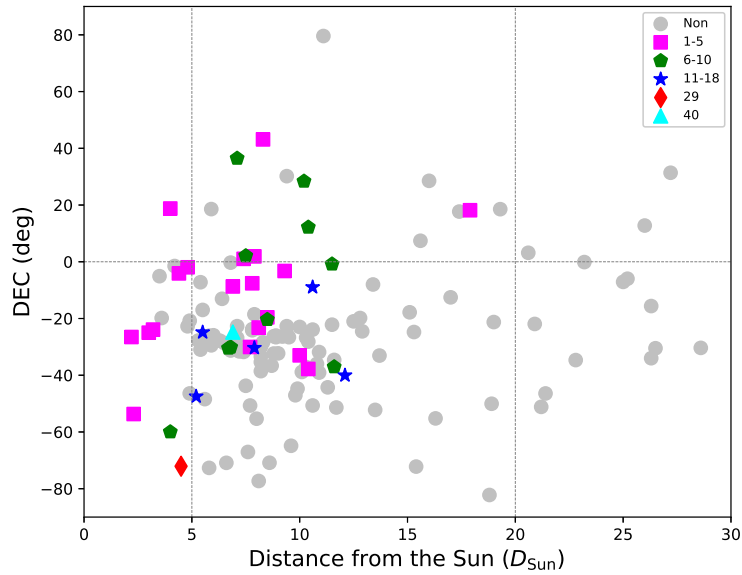}
    \caption{Distributions of GCs with and without pulsars in the distance vs. declination plane.
The different symbols with colours represent the
different numbers range of known pulsars in each GCs.
Dashed lines are for the declination 0$^{\circ}$ (horizontal) and distances from earth (vertical, for 5 kpc and 20 kpc), respectively.}
    \label{Fig04}
\end{figure*}

The fraction of GCs with known pulsars in all sky region is $\Gamma_{\rm all}$ = 36/141 $\approx$ $26\%$.
In the north and south sky,
pulsars were found in 9 out of 19 GCs ($\Gamma_{{\rm dec}>0^{\circ}} = 9/19 \approx 47 \%$)
and 27 out of 122 GCs ($\Gamma_{{\rm dec}<0^{\circ}} = 27/122 \approx 20 \%$), respectively (see Figure \ref{Fig04}).
Since 130 of the 157 GCs are located at declinations below 0$^{\circ}$,
radio telescopes covering more southern sky would have a significantly greater prospect of discovering more GC pulsars.
Nine of 15 GCs with $D_{\rm Sun} \leq$ 5 kpc have known pulsars,
while 27 of 126 with $D_{\rm Sun} \geq$ 5 kpc and $D_{\rm Sun} \leq$ 17.9 kpc have known pulsars.
The corresponding proportions are $\Gamma_{D_{\rm Sun}\leq 5}$ = 9/15 = $60\%$ and $\Gamma_{D_{\rm Sun} > 5}$ = 27/126 = $21\%$, respectively.
Thus,
higher sensitivity would still result in more GC pulsar discoveries.

\section{Dispersion measures and dispersion measure differences}
\label{DM and delDM}
For blind pulsar search, the time cost directly relates to the range of dispersion measure (DM).
Once the DM of any pulsars in a GC is known,
the search time will be largely reduced, since the DM values of these pulsars are similar.

\begin{figure*}
\centering
    \includegraphics[width=10cm]{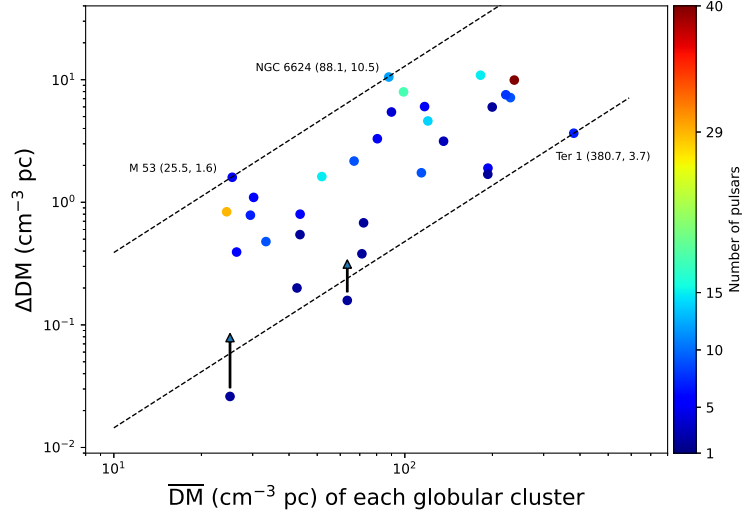}
    \caption{Relationship between the average of dispersion measures ($\overline{\rm DM}$ cm$^{-3}$pc) and the range of dispersion measure differences (${\rm DM}_{\rm max}-{\rm DM}_{\rm min}$) of pulsars in 31 GCs with known pulsars.
   The two clusters with arrows are M30 (25.08, 0.03) and NGC6652 (63.43, 0.16) respectively. These two clusters currently have 2 known pulsars, and the dispersion measure difference may increase if additional new pulsars are detected.}
    \label{Fig05}
\end{figure*}

The different column density of interstellar medium along the line of sight lead to different DM values of pulsars in the same GC.
It is not easy to tell whether a pulsar is a member of the GC when it is the only know pulsar.
For GCs host two or more pulsars, their close DM values can be the proof.
In the 36 GCs with known pulsars, 31 GCs host at least two pulsars.
Among these GCs,
the three with the largest $\Delta$DM values are 10.9 cm$^{-3}$ pc for NGC 6517,
10.5 cm$^{-3}$ pc for NGC 6624 and 9.9 cm$^{-3}$ pc for Terzan 5, respectively.
Figure \ref{Fig05} shows the relation between the average dispersion measures ($\overline{\rm DM}$)
and the differences of the dispersion measures ($\Delta {\rm DM} \equiv {\rm DM}_{\rm max}-{\rm DM}_{\rm min}$ in each GC)
for the pulsars in these 31 GCs.
It seems that the relation between $\Delta$DM and $\overline{\rm DM}$ is a power law.
With M53 and NGC6624, the equation for the maximum $\Delta$DM values is

\begin{equation}\label{fomu5}
\lg\Delta {\rm DM} = 1.52 \times \lg \overline{\rm DM} -1.93.
\end{equation}

For the lower boundary of the maximum $\Delta$DM value, only the data from Terzan 1 pulsars are used.
This is due to that M30 and NGC6652 (the two points below the line) only have 2 pulsars and $\Delta$DM may be increased with new discoveries.
Assuming the same slope as Equ. \ref{fomu5}, the equation for the lower boundary is

\begin{equation}\label{fomul2_1}
\lg\Delta {\rm DM} = 1.52 \times \lg \overline{\rm DM} -3.36.
\end{equation}

Thus, the pulsars in a GCs with higher $\overline{\rm DM}$ tend to distribute in a wider DM range.
This is similar to previous studies (e.g., \citealt{2005ApJ...621..959F}).
However, the uplimit of DM range can still be too wide when searching for pulsars in high DM GCs.

The uplimit can also be used as a proof as a pulsar is associated with the GC or or not.
For example, Terzan 5B (PSR J17482-2444, 205  cm$^{-3}$ pc, \citealt{1990Natur.347..650L}) was proved to be a foreground pulsar (\citealt{2005Sci...307..892R}).
The uplimit of DM difference of Terzan 5 from Equ. \ref{fomu5} is 48  cm$^{-3}$ pc ($\overline{\rm DM}
\pm $ 24 cm$^{-3}$ pc),
Terzan 5 is out of the limit.

\section{Conclusion}           
\label{sect:conclusion}

Till November 2022,
267 pulsars have been detected in 36 GCs. In these pulsars, 147 are in binary systems and 248 are millisecond pulsars with spin periods less than 30 ms.
Among the GCs pulsars, millisecond pulsars and binary pulsar systems account for 93 and 55$\%$, respectively.

We conclude that,

1. The sensitivity can limit the finding of new GC puslars. GC pulsar discoveries are mainly from the large radio telescopes.
Based on the detection rate, we suggest that FAST may find GC pulsars as many as it can, while MeerKAT should have many GC pulsars in the near future.

2. With the DM ranges of GC pulsars, we give the equations to estimate the DM ranges of pulsars in a GC.
The equations are

\begin{equation}\label{fomul2}
\lg\Delta {\rm DM} = 1.52 \times \lg \overline{\rm DM} -1.93  \nonumber
\end{equation}

for the upper bound and

\begin{equation}\label{fomul2_11}
\lg\Delta {\rm DM} = 1.52 \times \lg \overline{\rm DM} -3.36  \nonumber
\end{equation}

for the lower boundary.
The upper bound can also be used to judge that if a pulsar is a member of the GC or not.

3. For current spatial distributions of GCs and pulsars in them,
we suggest that there are higher changes to find new pulsars in GCs in the southern sky. In order to find more pulsars,
large radio telescopes in southern hemisphere should be used.

\normalem
\begin{acknowledgements}

This work is supported by National SKA Program of China No. 2020SKA0120100,
National Nature Science Foundation of China (NSFC) under Grant No. 11963002, 11703047, 11773041, U2031119, 12173052, 12003047 and 12173053.
We also thank the fostering project of GuiZhou University with No. 201911, and Cultivation Project for FAST Scientific Payoff and Research Achievement of CAMS-CAS.
This work is also supported by the Specialized Research Fund for State Key Laboratories. 
\\
ZP is supported by the CAS "Light of West China" Program and the Youth Innovation Promotion Accosciation of CAS (id 2023064).
LQ is supported by the Youth Innovation Promotion Association of CAS (id.~2018075 and Y2022027), and the CAS "Light of West China" Program.
This work made use of data from the Five-hundred-meter Aperture Spherical radio Telescope (FAST).
FAST is a Chinese national mega-science facility,
built and operated by the National Astronomical Observatories, Chinese Academy of Sciences (NAOC).
We appreciate all the people from the FAST group for their support and assistance during the observations. BL is supported by Guizhou Provincial Science and Technology Projects No. ZK[2023] 039 and [2023] 352.

\end{acknowledgements}

\bibliography{psrrefs}

\begin{table*}[htpb]
\centering
\caption{267 pulsars in 36 GCs. $N_{ALL}$, $N_{BP}$, $N_{EP}$, $N_{MSP}$, $N_{RB}$, $N_{BW}$, $N_{TBD}$ are all pulsar, binary pulsar, millisecond pulsar ($P < 30\ ms$),
eclipsing pulsar, redback pulsar,  black widow pulsar and to be defined pulsar in each GC.
$\Gamma$, the GCs stellar encounter rates catalog \citep{2013ApJ...766..136B}.
C: King-model central concentration, a 'c' denotes a core-collapsed cluster from (\citealt{1996AJ....112.1487H}, 2010 edition). M3C and NGC 6749B are likely real but need confirmation.
The pulsar data from Paulo¡¯s GC Pulsar Catalog (http://naic.edu/$\sim$pfreire/GCpsr.html, 2022-11 version).}
\label{tab1}
\begin{tabular}{cccccccccccc}
\hline
GC Name                   &  $N_{ALL}$             & $N_{BP}$    & $N_{EP}$       & $N_{MSP}$ &          $N_{RB}$ &  $N_{BW}$ &  $N_{TBD}$  &  $\Gamma$    &$C$ & $\frac{N_{MSP}}{N_{ALL}}$ ($\%$) & $\frac{N_{BP}}{N_{ALL}}$ ($\%$) \\
\hline
Terzan 5                     & 40                & 21             & 4                 & 38                 & 3       & 1           &       & 6800             & &97.50     & 52.50    \\
47 Tucanae (NGC 104)         & 29                & 19             & 7                 & 29                 & 3       & 6           &       & 1000            &    & 100.00    & 65.52    \\
$\omega$ Centauri (NGC 5139) & 18                & 5              & 1                 & 18                 & 0       & 1           &       & 90.4             &    & 100.00    & 27.78    \\
NGC 6517                     & 15                & 3              & 0                 & 14                 & 0       & 0           &       & 338            &    & 93.33     & 20.00    \\
NGC 1851                     & 15                & 9              & 0                 & 14                 & 0       & 0           &       & 1530            &    & 93.33     & 60.00    \\
M28 (NGC 6626)               & 14                & 10             & 2                 & 13                 & 2       & 5           &       & 648            &    & 92.86     & 71.43    \\
NGC 6624                     & 12                & 2              & 1                 & 9                  & 1       & 0           &       &1150             & c  & 75.00     & 16.67    \\
M62 (NGC 6266)               & 9                 & 9              & 2                 & 9                  & 1       & 2           &       &1670             & c: & 100.00    & 100.00   \\
NGC 6441                     & 9                 & 3              & 0                 & 7                  & 0       & 0           &       &2300             &    & 77.78     & 33.33    \\
NGC 6752                     & 9                 & 1              & 0                 & 9                  & 0       & 0           &       &401             & c  & 100.00    & 11.11    \\
M15 (NGC 7078)               & 9                 & 1              & 0                 & 5                  & 0       & 0           &       &4510             & c  & 55.56     & 11.11    \\
NGC 6440                     & 8                 & 4              & 1                 & 7                  & 1       & 1           &       &1400             &    & 87.50     & 50.00    \\
M5 (NGC 5904)                & 7                 & 6              & 1                 & 7                  & 0       & 1           &       &164             &    & 100.00    & 85.71    \\
Terzan 1                     & 7                 & 0              & 0                 & 6                  & 0       & 0           &       &0.292             & c  & 85.71     & 0.00     \\
M3 (NGC 5272)                & 6                 & 6              & 0                 & 6                  & 0       & 1           & 1     &194             &    & 100.00    & 100.00   \\
M13 (NGC 6205)               & 6                 & 4              & 1                 & 6                  & 0       & 1           &       &68.9             &    & 100.00    & 66.67    \\
M2 (NGC 7089)                & 6                 & 6              & 0                 & 6                  & 0       & 0           &       &518             &    & 100.00    & 100.00   \\
M53 (NGC 5024)               & 5                 & 4              & 0                 & 4                  & 0       & 0           &       &35.4             &    & 80.00     & 80.00    \\
M14 (NGC 6402)               & 5                 & 5              & 2                 & 5                  & 2       & 1           &       &124             &    & 100.00    & 100.00   \\
NGC 6522                     & 5                 & 0              & 0                 & 5                  & 0       & 0           &       &363             & c  & 100.00    & 0.00     \\
M71 (NGC 6838)               & 5                 & 5              & 1                 & 3                  & 0       & 2           &       &2.05             &    & 60.00     & 100.00   \\
M22 (NGC 6656)               & 4                 & 2              & 0                 & 4                  & 0       & 1           &       &77.5             &    & 100.00    & 50.00    \\
NGC 6544                     & 3                 & 3              & 0                 & 2                  & 0       & 1           &       &111             & c: & 100.00    & 100.00   \\
M12 (NGC 6218)               & 2                 & 2              & 0                 & 2                  & 0       & 0           &       &13             &    & 100.00    & 100.00   \\
M10 (NGC 6254)               & 2                 & 2              & 0                 & 2                  & 0       & 0           &       &31.4             &    & 100.00    & 100.00   \\
NGC 6342                     & 2                 & 1              & 1                 & 1                  & 0       & 0           &       &44.8             & c  & 50.00     & 50.00    \\
NGC 6397                     & 2                 & 2              & 2                 & 2                  & 2       & 0           &       &84.1             & c  & 100.00    & 100.00   \\
NGC 6652                     & 2                 & 2              & 0                 & 2                  & 0       & 0           &       &700             &    & 100.00    & 100.00   \\
NGC 6749                     & 2                 & 2              & 0                 & 2                  & 0       & 0           & 1     &38.5             &    & 100.00    & 100.00   \\
NGC 6760                     & 2                 & 1              & 0                 & 2                  & 0       & 1           &       &56.9             &    & 100.00    & 50.00    \\
M30 (NGC 7099)               & 2                 & 2              & 1                 & 2                  & 1       & 0           &       &324             & c  & 100.00    & 100.00   \\
NGC 5986                     & 1                 & 1              & 0                 & 1                  & 0       & 0           &       &61.9             &    & 100.00    & 100.00   \\
M4 (NGC 6121)                & 1                 & 1              & 0                 & 1                  & 0       & 0           &       &26.9             &    & 100.00    & 100.00   \\
M92 (NGC 6341)               & 1                 & 1              & 1                 & 1                  & 1       & 0           &       &270             &    & 100.00    & 100.00   \\
NGC 6539                     & 1                 & 1              & 0                 & 1                  & 0       & 0           &       &42.1             &    & 100.00    & 100.00   \\
NGC 6712                     & 1                 & 1              & 0                 & 1                  & 0       & 1           &       &30.8             &    & 100.00    & 100.00   \\
ALL                          & 267               & 147            & 28                & 248                & 17      & 26          & 2     &                &      & 92.88     & 55.05   \\
\hline
\end{tabular}
\end{table*}

\end{document}